\documentclass[conference]{IEEEtran}
\IEEEoverridecommandlockouts
\usepackage{cite}
\usepackage{amsmath,amssymb,amsfonts}

\usepackage{graphicx}
\usepackage{textcomp}
\usepackage{xcolor}
\usepackage{hyperref}
\def\BibTeX{{\rm B\kern-.05em{\sc i\kern-.025em b}\kern-.08em
    T\kern-.1667em\lower.7ex\hbox{E}\kern-.125emX}}

\usepackage{tikz}
\usepackage{booktabs}
\usepackage{algorithm}
\usepackage{algpseudocode}
\usepackage{pgfplots}
\usepackage{subcaption}
\usepackage{pgfplotstable}
\usepackage{colortbl}
\usepackage{siunitx}    
\usepackage{multirow}   
\usepackage{makecell}
\usepackage{adjustbox}
\usepackage{xspace}
\usepackage{enumitem}

\usepackage{amsthm}

\newcommand{\eg}{e.g.\xspace}
\usepgfplotslibrary{groupplots}
\usetikzlibrary{patterns}
\pgfplotsset{compat=1.18}
\newcommand*\circled[1]{\tikz[baseline=(char.base)]{
    \node[shape=circle,draw,fill=black,text=white,inner sep=1pt] (char) {#1};}}

\newcommand{\PHB}[1]{\noindent\textbf{#1}\hspace{.5em}} 
\newcommand{\PHM}[1]{\vspace{.4em} \noindent\textbf{#1}\hspace{.5em}} 
\newcommand{\YPHM}[1]{%
  \par\smallskip
  \noindent\textbf{#1}\hspace{.5em}%
}

\newcommand{\revise}[1]{\textcolor{black}{#1}}

\newcommand{\arachne}{\texttt{Arachne}}

\begin{document}

\date{}

\title{Arachne: Orchestrating Cascades for Efficient Text-to-Video Model Training}


\author{\IEEEauthorblockN{Peng Yu{$^{2,1}$}, Yuankai Fan{$^{1,*}$}, Yang Qiu{$^{1}$}, Tian Li{$^{1}$}, Bihuan Chen{$^{2}$}, Yin Chen{$^{1}$}, Qizhen Weng{$^{1,*}$}}
\IEEEauthorblockA{
  {$^1$}\textit{Institute of Artificial Intelligence (TeleAI)}, \textit{China Telecom}, Shanghai, China; \\
  \{fanyk1, lit117, cheny304, wengqzh\}@chinatelecom.cn, qiuy@hdu.edu.cn\\
  {$^2$}\textit{College of Computer Science and Artificial Intelligence}, \textit{Fudan University}, Shanghai, China; \\
  pyu25@m.fudan.edu.cn, bhchen@fudan.edu.cn}
  \thanks{$^*$Corresponding authors: Yuankai Fan, Qizhen Weng.}
}

\maketitle

\begin{abstract}
The rising demand for AI-generated videos is fueled by advances in large-scale Text-to-Video (T2V) models, trained on extensive datasets of video clips spanning diverse resolutions and durations. To address this data heterogeneity, current training methods often use a bucketing strategy that groups samples into discrete buckets for efficiency. However, this approach struggles to scale with compute and data volumes under static parallelism schemes, such as data and sequence parallelism, leading to significant workload imbalances and hardware under-utilization.

In this paper, we present Arachne, a novel training framework for efficient T2V model training at scale. Arachne decomposes the training process into fine-grained computational units, called \textit{cascades}, orchestrating their distributed execution and synchronization across the cluster through coordinated spatial and temporal optimization. Our comprehensive evaluation demonstrates that Arachne reduces iteration time by up to 65\% over leading frameworks, exhibiting a positive scaling trend where its performance advantages amplify as training scale grows.

\end{abstract}

\section{Introduction}

Video synthesis with deep generative models is transforming diverse domains, including media content creation~\cite{make-a-video23, blattmann2023stable, Bacher20, moviegen2025}, interactive video games~\cite{yu2025gamefactorycreatingnewgames, Valevski25, che2024gamegenx}, and world simulation~\cite{world-model18, WorldSimBench24, feng2024matrix}.  Among these, Text-to-Video (T2V) models emerge as the leading and rapidly advancing frontier, driven by advances in generative algorithms~\cite{ho2020denoising, score_based, latent_diffusion, lipman2022flow} , scalable model architectures~\cite{scalable-diffusion, Latte25, Photorealistic24}, and unprecedented availability of compute and data resources~\cite{MiraData24, OpenVid25, wang2024koala36mlargescalevideodataset}.

\begin{figure}[t]
  \centering

  \begin{subfigure}{\linewidth}
    \centering
    \includegraphics[width=\linewidth]{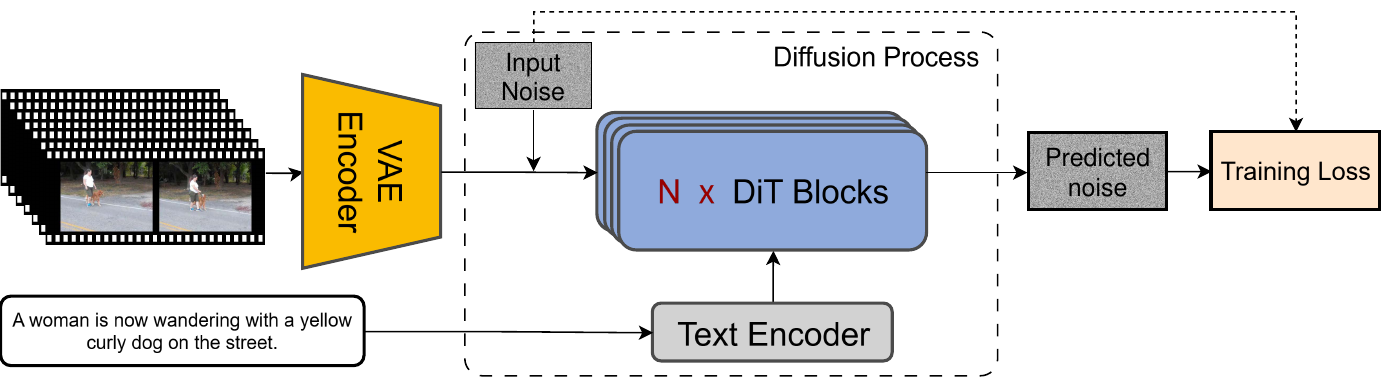}
    \caption{Training workflow of a typical T2V model, consisting of a \textit{VAE}, a \textit{text encoder}, and multiple \textit{DiT blocks}.}
    \label{fig:t2v_model}
  \end{subfigure}

  \vspace{1mm}

  \begin{subfigure}{\linewidth}
    \centering
    \includegraphics[width=\linewidth]{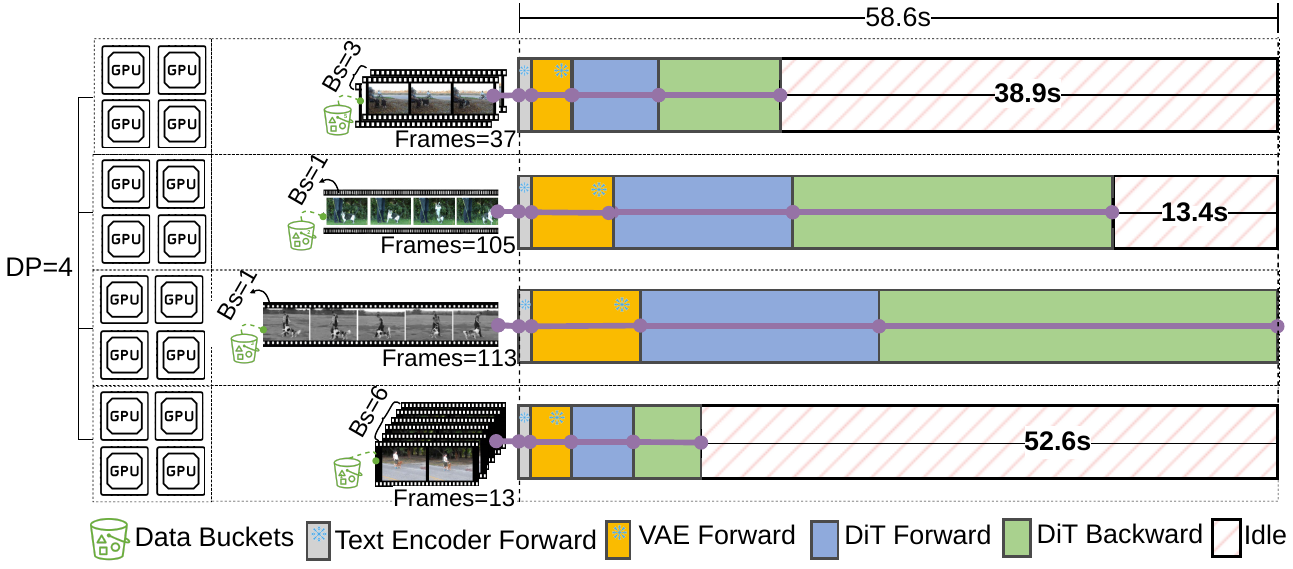}
    \caption{Illustration of a T2V model training step under the default bucketing (4 frame-length buckets) and parallelism configurations (\texttt{DP=4}, \texttt{SP=4}), showing \textit{significant GPU idle time (hatched regions)} induced by the gradient synchronization barrier following the DiT backward pass.}
    \label{fig:load_imbalance}
  \end{subfigure}

\end{figure}

Current state-of-the-art T2V models~\cite{OpenAI2024Sora, moviegen2025, Kong2024HunyuanVideo, Wang2025Wan, Yang2025CogVideoX}, have converged on the latent diffusion paradigm~\cite{latent_diffusion} as their architectural blueprint. As illustrated in~\autoref{fig:t2v_model}, this paradigm involves a Variational Autoencoder (VAE)~\cite{VAE} first compressing video into a latent space, where a Diffusion Transformer (DiT)~\cite{scalable-diffusion}, guided by a text encoder, is then trained to predict artificially added noise. Here, the success of this paradigm in generating high-quality video relies heavily on training with massive datasets~\cite{moviegen2025, Zheng2024OpenSora, Kong2024HunyuanVideo}.



Training on these massive datasets is typically accelerated using data parallelism (DP)~\cite{Li2020PyTorchDP}, where the workload is evenly distributed across multiple devices. However, due to the data's inherent heterogeneity, a local batch randomly sampled for each DP rank will contain a mix of videos with varying durations and resolutions, making it difficult to efficiently organize the data into uniform batches for training. The widely adopted solution is the \textit{bucketing strategy}~\cite{Zheng2024OpenSora, Kong2024HunyuanVideo, ma2025stepvideo}, which pre-groups videos by similar resolutions and durations so that each DP rank can sample a local batch from a single bucket. In addition, for buckets containing long sequences that exceed single-device memory, sequence parallelism (SP)~\cite{ulysess, korthikanti2023reducing, Li2024DistFlash, Li2023Sequence} is utilized to partition a single data sequence across a dedicated group of devices.

\definecolor{vae}{RGB}{250, 186, 3}
\definecolor{dit}{RGB}{143, 170, 220}

\PHM{Challenges.}However, while bucketing ensures uniformity within each local batch, it still leaves a considerable workload imbalance across DP ranks. The strategy attempts to mitigate this issue by incorporating adaptive batch sizing\footnote{Buckets with shorter sequences are assigned larger batch sizes to equalize the total tokens (i.e., $\mathrm{batch\_size} \times \mathrm{sequence\_length}$) processed per rank, subject to GPU memory constraints}, \revise{but such adjustments are fundamentally constrained by GPU memory limits and the rigidity of static parallelism.} This is clearly illustrated by the example in~\autoref{fig:load_imbalance}. In a representative training step in the T2V model HunyuanVideo~\cite{Kong2024HunyuanVideo}, executed on four \texttt{DP=4} ranks, each with an \texttt{SP=4} degree, four distinct batches from separate buckets are fed into the SP groups, with sequence lengths varying from 13 to 113 frames. Due to the static parallelism configuration applied to all groups, the rank processing the longest sequence (113 frames) takes 58.6 seconds, whereas the one with the shortest (13 frames) completes in only 6.0 seconds, resulting in 52.6 seconds of idle time for the fastest rank. This example shows the substantial resource under-utilization inherent in current training schemes.

The challenge of mitigating inefficiencies from static parallelism under data heterogeneity has spurred the development of \textit{dynamic parallelism approaches} in the Large Language Model (LLM) field, primarily targeting the communication overhead of short text sequences~\cite{Wang2025FlexSP, Ge2024HotSwitch}. However, such a focus is mismatched with the realities of T2V generation, where the combination of significantly longer video sequences and compute-intensive DiT architectures makes computation, not communication, the defining performance bottleneck (more details can be seen in \S\ref{sec: observations}).

\PHM{Proposed Method.} This paper proposes \arachne{}, a novel training framework tailored for T2V model training. At its core, \arachne{} introduces a unified strategy that integrates fine-grained workload decomposition with dynamic spatio-temporal orchestration. It first breaks down the training process into minimal schedulable units known as \textit{cascades}, each a self-contained task defined by a bespoke configuration that specifies its target model components, input data, parallelization strategy, and scheduling parameters. This decomposition empowers \arachne{} to execute these cascades with flexibility and efficiency via spatio-temporal orchestration.



Specifically, \arachne{} introduces three components that work collaboratively to optimize T2V training. \circled{1}\ \textit{Cascade-level parallelism planner} performs temporal optimization by decomposing the training iteration into a Directed Acyclic Graph (DAG) of dependent cascades, co-optimizing the launch time and parallel configuration for each to minimize the theoretical makespan on abstract, homogeneous hardware. \circled{2}\ \textit{Topology-aware resource mapper} then performs spatial optimization by taking the planner's logical, topology-agnostic plan and, considering the physical network topology, mapping each cascade onto a specific set of GPUs to create a communication-efficient physical execution plan.  Finally, ~\circled{3}\ \revise{\textit{Runtime executor}} enacts the concrete spatio-temporal plan by coordinating data handoff across dependent cascades and regulating gradient accumulation within parallelism groups, thereby ensuring both correctness and efficiency.

We evaluate \arachne{}'s performance on three large-scale T2V models, including Wan2.1~\cite{Wang2025Wan}, CogVideoX~\cite{Yang2025CogVideoX}, and HunyuanVideo~\cite{Kong2024HunyuanVideo}, and compare it with two leading training frameworks—Megatron-LM~\cite{Shoeybi2019MegatronLM} and DeepSpeed~\cite{Rajbhandari2020ZeRO}—as well as FlexSP~\cite{Wang2025FlexSP}, a representative training framework tailored for variable-length corpora. Our experimental results show that, under realistic T2V training scenarios, \arachne{} achieves significant end-to-end speedups, reducing average iteration time by up to 65\%, 59\%, and 35\% compared to these frameworks, respectively. To further evaluate scalability, we examine \arachne{} across diverse model architectures, workload heterogeneities, and cluster sizes, which reveals a clear \textbf{scaling trend}: as training scale increases, the speedup achieved by \arachne{} grows correspondingly, reflecting the ``scaling-law'' behavior observed in model performance under realistic large-scale training scenarios.

\PHM{Contributions.}We make the following contributions:
\begin{enumerate}[leftmargin=*]
\setlength\itemsep{0.2em}
\item We present \arachne{}, a novel training framework for efficient large-scale T2V model training. Its design is informed by a systematic analysis of T2V's unique workload characteristics, revealing performance bottlenecks distinct from those encountered in the LLM domain.

\item We propose a cascade-level spatio-temporal orchestration paradigm that mitigates workload imbalance arising from data heterogeneity. This paradigm temporally optimizes the launch time and parallel configuration for each fine-grained cascade, while spatially mapping it to communication-efficient GPU placements.

\item Through comprehensive experiments, we show that \arachne{} achieves significant and consistent speedups over leading baselines, demonstrating both effectiveness and robust scalability across diverse T2V models, workload heterogeneities, and cluster configurations.

\item We will open-source the full implementation of \arachne{} and the T2V training workloads used in our evaluation, fostering reproducibility and enabling further research.
\end{enumerate}


\section{Background} \label{sec: background}
This section provides an overview of the core architectures, training paradigms, and parallelization strategies used in modern T2V models.

\subsection{T2V Diffusion Models}
Modern T2V models typically follow a multi-module latent diffusion paradigm~\cite{latent_diffusion}, as illustrated in~\autoref{fig:t2v_model}, comprising a VAE for compression, a text encoder for conditioning, and a DiT backbone for denoising. In this paradigm, the VAE (often a 3D causal VAE~\cite{3d_casual_vae}) encodes raw video clips into compact spatio-temporal latent representations. Noise is added to these latents, and under text conditioning, the DiT is trained to predict the added noise in DDPM~\cite{ho2020denoising} or the velocity toward clean latents in Flow Matching~\cite{lipman2022flow, ma2024sit}.

\subsection{T2V Training Strategies} \label{subsec:t2v_training_strategies}

Training T2V models on large-scale, heterogeneous datasets typically relies on two core strategies: multi-stage curriculum learning to ensure stable convergence, and data bucketing to enable efficient batching.

\PHM{Curriculum Strategy.}To make training T2V models computationally tractable and stable, models follow a multi-stage, coarse-to-fine curriculum~\cite{Kong2024HunyuanVideo,Zheng2024OpenSora,chen2025goku}. Training starts on massive, low-resolution video-text data to establish text-video alignment, and then progressively fine-tunes on smaller, higher-quality datasets at increasing resolutions to improve visual fidelity and temporal coherence.

\PHM{Bucketing Strategy.} To batch heterogeneous videos, naive zero-padding incurs massive computational waste, while adapting sequence packing~\cite{sequence-packing} to 3D video tensors introduces extreme implementation complexity and inefficient attention masking. As a practical alternative, the bucketing strategy~\cite{Zheng2024OpenSora, Kong2024HunyuanVideo} pre-categorizes videos into discrete buckets based on fixed resolutions and durations, thereby guaranteeing strict intra-batch homogeneity during training.

\subsection{Parallelisms in T2V Training}\label{subsec:parallelisms_in_distributed_training}
Unlike LLMs that rely on Tensor and Pipeline Parallelism (TP/PP) to distribute massive parameters~\cite{Shoeybi2019MegatronLM,GPipe19}, T2V training is primarily constrained by the activation memory of long video sequences. Consequently, practitioners favor data-centric parallelism: Data Parallelism (DP) for throughput scaling, and Sequence/Context Parallelism (SP/CP) to shard long contexts~\cite{Zheng2024OpenSora,Fan2025Vchitect20,Wang2025Wan}.


\PHM{Data Parallelism (DP).}DP replicates the model on each device while sharding the training data. During each step, all devices process their local data in parallel before a global gradient synchronization ensures a consistent model update. This approach is often enhanced with optimizers like ZeRO~\cite{Rajbhandari2020ZeRO} that also partition optimizer states, gradients, and parameters to further reduce per-device memory.



\PHM{Sequence/Context Parallelism (SP/CP).}SP and CP shard long video contexts across devices to reduce activation memory. Ulysses-style SP uses \textit{head partitioning} by splitting attention heads across devices and relying on \texttt{All-to-All} communication to enable full-sequence self-attention for each head shard~\cite{ulysess}. In contrast, CP uses \textit{sequence partitioning} by sharding $Q/K/V$ along the sequence dimension and computing attention block-wise while circulating $K/V$ slices in a ring, which enables overlapping communication with computation~\cite{Brandon2023Striped,Liu2023RingAttention,lightseq,nvidia2024contextparallelism}.

\section{Observations} \label{sec: observations}
Existing distributed training systems, heavily optimized for LLMs, fall short for T2V generation, whose distinct sequence lengths and computation patterns create new bottlenecks. We highlight two key observations that expose this mismatch and motivate a T2V-aware system design.

\begin{figure}[t]
  \centering
  \includegraphics[width=0.8\linewidth]{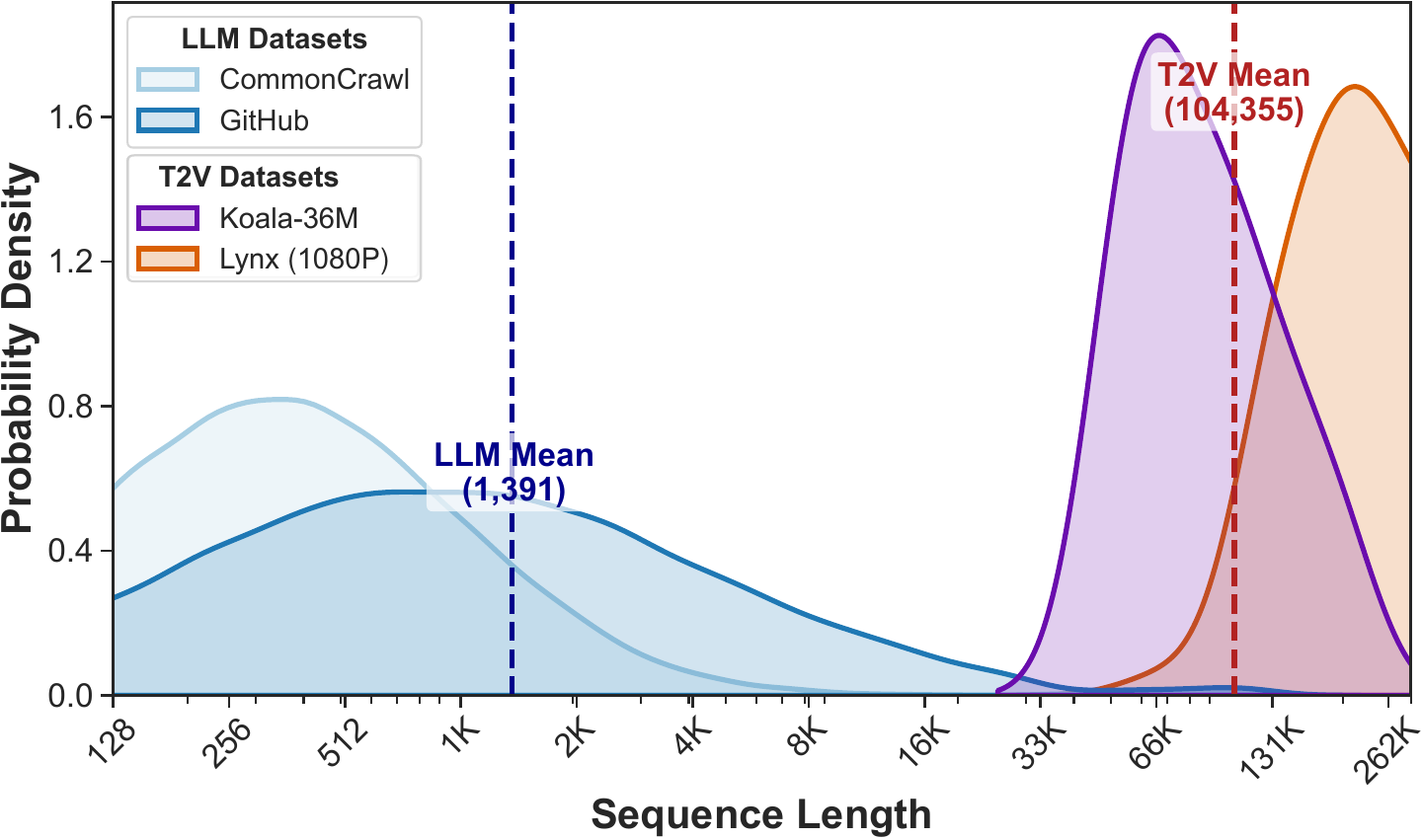}
  \caption{\textit{Sequence-length distributions} for two T2V datasets (Koala~\cite{wang2024koala36mlargescalevideodataset} and an internal $1080$p dataset called Lynx) and two LLM datasets (CommonCrawl and GitHub). The x-axis is shown in log scale for readability. The \textit{vertical dashed lines} mark the average sequence lengths for each domain.}
  \label{fig:seq_dist_comparison}
\end{figure}

\PHM{Observation 1: Right-shifted, concentrated, and discrete sequence-length distributions in video workloads.}
The primary distinction between T2V and LLM workloads lies in their sequence-length distributions, as depicted in~\autoref{fig:seq_dist_comparison}. Comparing T2V datasets with commonly used LLM corpora highlights three salient characteristics of video workloads that challenge conventional scheduling assumptions: they are \textit{right-shifted}, \textit{concentrated}, and \textit{discrete}.



\begin{itemize}[leftmargin=*]
 \item\textit{Right-shifted}. The entire distribution is shifted dramatically to the right as a direct consequence of the data modality; after VAE encoding, even short video clips generate thousands of tokens, making the shortest T2V sequences more computationally demanding than the vast majority of LLM sequences (\eg, the 99.5\% shorter than 8K tokens). 

 \item\textit{Concentrated}. Unlike the long-tail distribution of LLM workloads that spans several orders of magnitude (\eg, from 128 to 128K tokens), T2V workloads exhibit a high relative concentration. Despite a large absolute token variance, the dynamic range of T2V sequences (\eg, 33K to 262K) is limited to a small ratio ($\sim 8\times$). This clustering within a high-cost computational region leads to a sustained and predictable heavy load.
 \item\textit{Discrete}. Since sequence length increases in discrete increments determined by video frame counts, marginal variations in video duration can lead to substantial computational disparities. Such discontinuity is a major source of inefficiency that amplifies the load imbalance across pre-defined training buckets.
\end{itemize}

\begin{figure}[!h]
  \centering
\includegraphics[width=\linewidth]{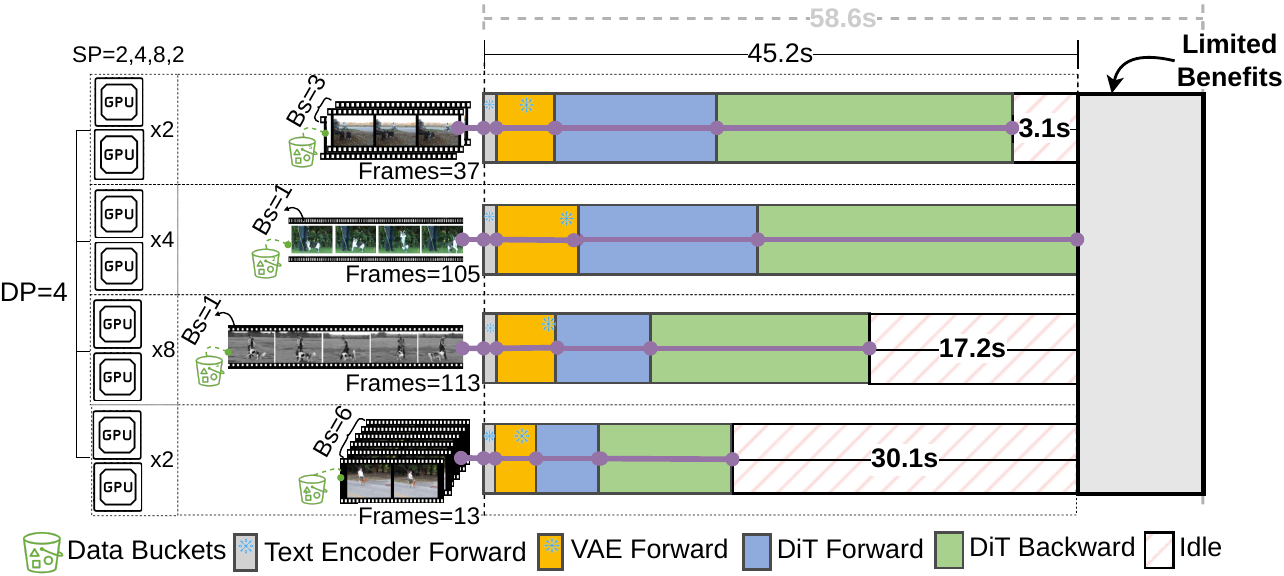}
  \caption{Training step corresponding to~\autoref{fig:load_imbalance}, with parallelism \textit{reconfigured} across buckets (\texttt{SP=8} for the $113$-frame bucket and \texttt{SP=2} for the $13$- and $37$-frame buckets).}
  \label{fig:dynamic_parallelism_flex}
\end{figure}

\begin{figure*}[t]
    \centering
    \includegraphics[width=\textwidth]{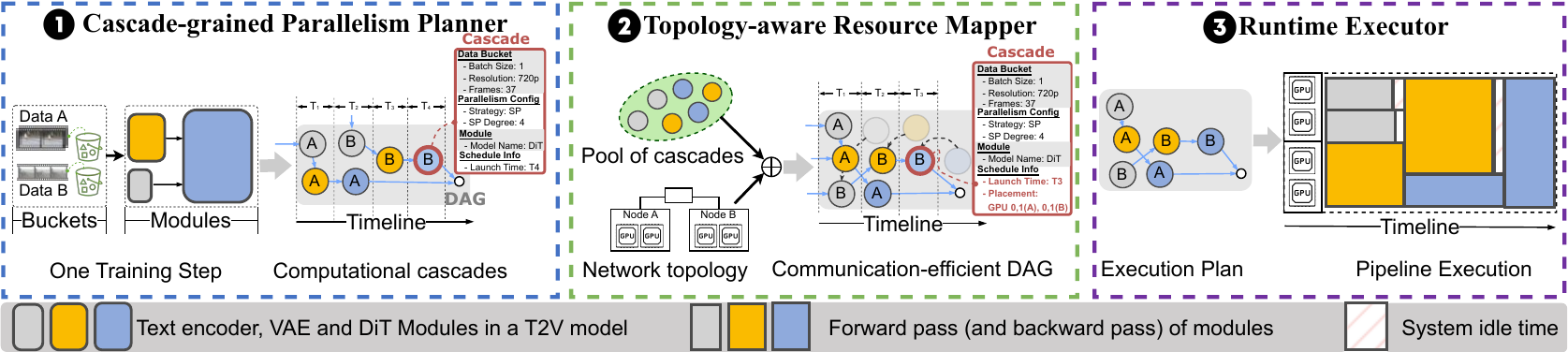}
    \caption{Overview of \arachne{}.}
    \label{fig:overview}
\end{figure*}

\YPHM{Observation 2: Computation as the bottleneck and the limits of LLM-style scheduling.}\label{subsec:llm-style-schedule}A second defining characteristic of the T2V workloads is their extreme computational intensity, stemming from the DiT architecture's full self-attention over a unified sequence of 3D patches of length $L$, which incurs quadratic ($O(L^2)$) complexity~\cite{tan2025dsv,zhang2025sta,zhang2025vsa}. When applied to the inherently long sequences described above, this operation becomes exceedingly \textit{compute-intensive}, emerging as the dominant performance bottleneck.

Clearly, this characteristic limits the applicability of many approaches derived from the LLM domain~\cite{Wang2025FlexSP, Ge2024HotSwitch}; existing frameworks are typically designed for text corpora with high variability in sequence length, where communication overhead across short sequences is the primary performance concern. However, these communication-oriented approaches are insufficient for compute-bound workloads observed in T2V.

\definecolor{cusgray}{RGB}{121,121,121}
\YPHM{Example 1.}Continuing the example in~\autoref{fig:load_imbalance}, we apply the communication-oriented dynamic parallelism method introduced in~\cite{Wang2025FlexSP}, which increases parallelism for the $113$-frame sequence to \texttt{SP=8} while reducing it for the $13$- and $37$-frame sequences to \texttt{SP=2}, respectively. \revise{ As shown in~\autoref{fig:dynamic_parallelism_flex}, such optimization yields only limited benefits (\textcolor{cusgray}{\textbf{gray region}}): while the overall latency is reduced from $58.6\text{s}$ to $45.2\text{s}$, the performance bottleneck merely shifts to the $105$-frame bucket, leaving substantial GPU idle times (i.e., $3.1\text{s}$, $17.2\text{s}$, and $30.1\text{s}$)}. This example shows that applying communication-oriented optimizations in LLMs is challenging for T2V training due to the computation-intensive nature of the workload.


\YPHM{Motivation for Our Approach.}Our analysis reveals that the interplay of heterogeneous video sequences and compute-intensive DiT architecture poses a unique training challenge for T2V models: one that existing parallelism approaches, including recent dynamic parallelism from the LLM field, are ill-equipped to solve. This inadequacy stems from two constraints that are inherent to all such approaches: (1) a \emph{static spatial assignment}, which binds each data sample to a fixed set of GPUs for its entire lifecycle, and (2) a \emph{strict temporal synchronization}, which mandates that all samples within an iteration launch simultaneously. Confronted with the profound heterogeneity of T2V data, these twin rigidities are the direct cause of the severe workload imbalance and resource under-utilization that plague current systems.

To overcome this rigidity, we introduce fine-grained spatio-temporal orchestration, a paradigm enabled by decomposing each sample's computation into minimal, schedulable units (i.e., cascades). Such fine-grained decomposition unlocks independent control across both dimensions: on the temporal axis, we can precisely schedule each cascade's launch time and assign it a bespoke parallel configuration to control its duration. On the spatial axis, we can map it onto a specific set of GPUs to create a communication-efficient physical layout. This flexible orchestration mitigates system idle time inherent in rigid schemes, thereby improving load balance and minimizing the end-to-end training makespan.

\section{System Design}\label{sec: method}
\subsection{Overview}
\arachne{} is a distributed training framework for T2V workloads by addressing the challenges outlined in \S\ref{subsec:llm-style-schedule}. A high-level view of \arachne{} is shown in~\autoref{fig:overview}, which illustrates its three main components: \textit{Cascade-level parallelism planner}, \textit{Topology-aware resource mapper}, and \textit{Runtime executor}.  We first provide an overview of each component and detail its design in the following subsections.

\PHM{Cascade-level parallelism planner.}The planner (~\autoref{fig:overview}-\circled{1}) performs temporal orchestration, directly dismantling the strict synchronization of traditional paradigms. It first decomposes each sample's execution into a DAG of fine-grained, schedulable cascades. It then formulates the scheduling of these cascades as an optimization problem to minimize the global theoretical makespan, generating a logical execution plan that co-optimizes the launch time and parallel configuration for each cascade.

\PHM{Topology-aware resource mapper.}While the planner generates a temporally optimal logical plan, it does so on a homogeneous hardware abstraction. The mapper (illustrated in~\autoref{fig:overview}-\circled{2}) addresses this by performing spatial orchestration to create a communication-efficient plan that breaks the spatial rigidity of fixed resource allocation. To achieve this, it refines the logical DAG via a two-level spatial arrangement: (1) At the intra-cascade level, it selects optimal physical GPUs to minimize a cascade's internal communication. (2) At the inter-cascade level, it co-locates dependent cascades to maximize data locality and reduce transfer costs.


\PHM{Runtime executor.}Runtime executor (illustrated in~\autoref{fig:overview}-\circled{3}) is the engine that enacts the complete spatio-temporal optimized execution plan crafted by the planner and mapper. It brings the optimized schedule to life by launching each cascade on its designated GPUs according to the precise timeline. Critically, the Executor manages the complex inter-cascade dependencies, handling state transitions (e.g., data handoffs) and synchronizations across different parallelism groups. To ensure correctness amidst this highly dynamic execution, it also implements a novel synchronization primitive for heterogeneous gradient accumulation, guaranteeing efficient and accurate gradient aggregation.


\subsection{Cascade-level Parallelism Planner}\label{sec: 4.2}
A fundamental question for \arachne{} is how to determine the optimal execution strategy for a given training step. To this end, Cascade-level parallelism planner decomposes the training process into a set of basic schedulable units, referred to as cascades, and formulates their execution as a global \textit{makespan minimization problem}, thereby deriving an optimized execution plan. In this section, we first provide a formal definition of the cascade and then present the corresponding optimization formulation.

\noindent\textsc{Definition} 1. \textbf{(Cascade)} \textit{A} cascade \textit{is the minimal, self-contained computational unit that can be directly executed within existing distributed training frameworks. Each} cascade \textit{defines the configuration required for distributed execution, including input data, module specifications, parallelization strategy, GPU allocation, and scheduling parameters.}

\PHM{Example 3.}Consider the case depicted in~\autoref{fig:overview}-\circled{1}. The cascade is characterized by: (1) \textit{input data}: one $37$-frame $720$p video clip; (2) \textit{model specification}: DiT; (3) \textit{parallelization strategy}: SP degree of $4$; (4) \textit{GPU allocation}: \texttt{GPU~0} and \texttt{1} on both node~A and node~B; and (5) \textit{launch time}: $T_3$.

\begingroup
\setlength{\abovedisplayskip}{5pt}
\setlength{\abovedisplayshortskip}{4pt}
\setlength{\belowdisplayskip}{5pt}
\setlength{\belowdisplayshortskip}{4pt}
\allowdisplaybreaks[4]

\vspace{1em}
\noindent\begin{minipage}{\columnwidth}
    \captionsetup{type=table}
    \caption{Nomenclature for the MILP Formulation.}
    \label{tab:milp_nomenclature}
    \footnotesize
    \setlength{\tabcolsep}{3pt}
    \centering
    \begin{tabular}{p{0.16\linewidth} p{0.74\linewidth}}
        \toprule
        \multicolumn{2}{l}{\textbf{Decision and Auxiliary Variables}} \\
        \midrule
        $s_c$              & Launch time of cascade $c$ \\
        $e_c$              & End time of cascade $c$ \\
        $x_{c,k}$          & Binary variable: 1 if cascade $c$ uses parallelism degree $k$, 0 otherwise \\
        $T$                & The makespan of the schedule (objective to be minimized) \\
        \midrule
        \multicolumn{2}{l}{\textbf{Sets and Indices}} \\
        \midrule
        $I$                & Set of bucket samples $i$ in the current batch \\
        $M$                & Set of model modules $m$ (\eg, VAE) \\
        $C$                & Set of all cascades $c$, where each $c$ corresponds to a pair $(i,m)$ \\
        $K$                & Set of all available parallelism degrees $k$ \\
        $A_t$              & Set of active cascades at time $t$, where $A_t = \{c \in C \mid s_c \le t < e_c\}$ \\
        $P_c$              & Set of direct predecessors of cascade $c$ \\
        \midrule
        \multicolumn{2}{l}{\textbf{Parameters}} \\
        \midrule
        $L_{c,k}$          & Latency of cascade $c$ with parallelism degree $k$ \\
        $\mathrm{Mem}_{c,k}$ & Peak memory requirement of cascade $c$ with parallelism degree $k$ \\
        $N$                & Total number of available GPUs \\
        $W$                & Maximum memory capacity of a single GPU \\
        \bottomrule
    \end{tabular}
\end{minipage}

\vspace{1em}
\begin{align}
    \text{Minimize}\quad & T \label{eq:objective} \\
    \text{subject to:} \nonumber \\[-0.2em]
    & e_{c} = \sum_{k \in K} (L_{c,k} \cdot x_{c,k}) + s_{c}, \quad \forall c \in C \label{eq:completion_time} \\
    & T \ge e_{c}, \quad \forall c \in C \label{eq:makespan_def} \\
    \displaybreak[3]
    & s_{c} \ge e_{p}, \quad \forall p \in P_c,\ \forall c \in C \label{eq:precedence} \\
    & \sum_{k \in K} x_{c,k} = 1, \quad \forall c \in C \label{eq:sp_uniqueness} \\
    & \sum_{k \in K} (\text{Mem}_{c,k} \cdot x_{c,k}) \le W, \quad \forall c \in C \label{eq:memory_constraint} \\
    & \sum_{c \in A_t} \sum_{k \in K} k \cdot x_{c,k} \le N, \quad \forall t \ge 0 \label{eq:resource_conceptual}
\end{align}
\endgroup

\PHM{Problem Formulation.}\label{sec:problem_formulation}
We formalize cascade scheduling as a makespan minimization problem, aiming to determine \textit{when} each cascade executes and \textit{how} much parallelism it uses to minimize overall execution time. To focus on the intrinsic scheduling complexity, we treat GPUs as a homogeneous resource pool and intentionally abstract away concrete GPU placement; the resulting formulation therefore captures a logical scheduling optimum that jointly optimizes execution order and parallelism under dependency, memory, and aggregate resource constraints.

The primary input is a dependency graph constructed from each training batch, where nodes represent cascades and edges encode data dependencies. Based on this abstraction, we model the scheduling problem as a \textit{Mixed-Integer Linear Programming} (MILP) formulation, with notation summarized in~\autoref{tab:milp_nomenclature}. The objective is to minimize the total makespan ($T$) by jointly optimizing each cascade’s launch time $s_c$ and parallel configuration $x_{c,k}$. The formulation enforces fundamental constraints on unique parallelism selection, precedence ordering, and per-GPU memory limits (Eqs.~\eqref{eq:precedence}–\eqref{eq:memory_constraint}), as well as an event-based resource constraint that bounds the number of concurrently active GPUs (Eq.~\eqref{eq:resource_conceptual}).

The efficacy of our MILP formulation (\S\ref{sec:problem_formulation}) requires accurate apriori performance estimates. To this end, we developed two cost models\footnote{We exclude the cost model for the text encoder in the T2V model, as its contribution to the total makespan is negligible compared to the dominant bottleneck of processing long video sequences. This is due to the typically short text sequences in T2V training (\eg under 256 tokens).} to predict the execution latency $L_{c,k}$ and peak GPU memory $\text{Mem}_{c,k}$ for each cascade $c$ with any potential parallelism degree $k$.

\PHM{Cost Model for VAE.} The VAE's 3D convolutional layers generate prohibitively large activation tensors when processing high-resolution or long-duration videos~\cite{Li2025WFVAE,Cheng2025LeanVAE}. To prevent out-of-memory (OOM) errors, the VAE distributes independent volumetric sub-tiles across GPUs. While this tile-level parallelism ensures memory efficiency, it severely complicates performance prediction. We resolve this via a hierarchical model synergizing lightweight simulation with machine learning. To estimate end-to-end latency for a parallelism degree $k$, a top-level simulator distributes standardized ``base tiles'' across workers round-robin, bounding the total latency by the makespan of the bottleneck GPU. Crucially, individual base-tile latencies are predicted via a two-stage hybrid approach: an analytical baseline (linear regression on tile volume) refined by an XGBoost~\cite{chen2016xgboost} model, which fits residuals using one-hot encoded spatial dimensions (\eg, $256 \times 256$) to capture shape-dependent hardware execution efficiencies.

\PHM{Cost Model for DiT.} For the Transformer-based DiT module, the cost model provides peak memory and latency estimates essential for the planner to derive memory-feasible, makespan-minimizing schedules. We formulate these estimates using batch size $B$ and sequence length $S$.

To prevent OOM errors, the per-GPU peak memory ($\mathrm{Mem}_{\mathrm{DiT}}$) accounts for a fixed model-state component ($\mathrm{Mem}_{\mathrm{states}}$)---sharded according to the device count $N$ and ZeRO stage---and a variable activation component. This activation memory is distributed across $k$ GPUs based on the total token count ($B \cdot S$) and the empirically profiled per-token memory ($\mathrm{Mem}_{\mathrm{token}}$):
\begin{equation}
\mathrm{Mem}_{\mathrm{DiT}}(B, S, k) = \mathrm{Mem}_{\mathrm{states}} + \frac{B \cdot S \cdot \mathrm{Mem}_{\mathrm{token}}}{k}
\label{eq:dit_mem_final}
\end{equation}

The latency model ($L_{\mathrm{DiT}}$) decomposes execution time into parallelizable computation and communication. To reflect on the DiT architecture, it utilizes regression coefficients $\alpha_1$ and $\alpha_2$ to model linear operations (\eg, MLPs) via $\alpha_1(B \cdot S)$, while capturing the $\mathcal{O}(S^2)$ complexity of the self-attention mechanism via $\alpha_2(B \cdot S^2)$. The unified formulation is expressed as:
\begin{equation}
L_{\mathrm{DiT}}(B, S, k) = \frac{\alpha_1(B \cdot S) + \alpha_2(B \cdot S^2)}{k} + \frac{V_{\mathrm{comm}}}{\mathrm{BW}_{\mathrm{eff}}(k)}
\label{eq:dit_cost_final}
\end{equation}
where $V_{\mathrm{comm}}$ is the communication volume, and $\mathrm{BW}_{\mathrm{eff}}(k)$ is the effective bandwidth determined by the parallelism degree $k$ and its corresponding intra- or inter-node characteristics.

\begin{algorithm}[ht]
\caption{Genetic Algorithm for Cascade Scheduling}
\begin{algorithmic}[1]
\small
\State \textbf{Input:} cascades $C$, cost model $\mathcal{M}$, and number of GPUs $N$

\State \textbf{Hyper-parameters:} population size $p$, generations $G$, mutation rate $\mu$, and elitism size $\epsilon$
\State \textbf{Output:} best makespan $T^*$ and best schedule $S^*$

\State $T^* \gets \infty$, $S^* \gets \text{None}$ \Comment{\textcolor{blue}{Initialize the best-found solution}}

\State $\mathcal{P} \gets$ \textsc{InitializePopulation}($C, \mathcal{M}, p$) \Comment{\textcolor{blue}{Initialize a population $\mathcal{P}$ of $p$ chromosomes}} 

\For{$g = 1 \to G$}
    \State $\{S_i\}, \{T_i\} \gets$ \textsc{EvaluateFitness}($\mathcal{P}, C, \mathcal{M}, N$) \Comment{\textcolor{blue}{Find a short makespan}} 
    
    \State $T_{gen}, S_{gen} \gets$ \textsc{FindBestInGeneration}($\{S_i\}, \{T_i\}$)
    
    \If{$T_{gen} < T^*$} \Comment{\textcolor{blue}{Update the global best solution}}
        \State $T^* \gets T_{gen}$
        \State $S^* \gets S_{gen}$
    \EndIf
    
    \State $\mathcal{P} \gets$ \textsc{EvolveNextGeneration}($\mathcal{P}, \{T_i\}, \mu, \epsilon$)\Comment{\textcolor{blue}{Via elitism, crossover, and mutation}}
\EndFor

\State \Return $T^*, S^*$ \Comment{\textcolor{blue}{Return the best schedule}}
\end{algorithmic}
\label{alg:ga_solver}
\end{algorithm}

\definecolor{line}{RGB}{119, 30, 126}
\PHM{Genetic Algorithm for Heuristic Cascade Scheduling.}\label{sec:planner_algorithm}
\revise{While the MILP formulation in~\S\ref{sec:problem_formulation} defines the theoretical optimum, the multi-module architecture of T2V workloads (\eg, Text Encoder, VAE, DiT) inherently decomposes each sample into multiple distinct cascades, across which the joint optimization of parallelism configurations and timeline scheduling induces a combinatorial explosion. We therefore adopt a \textit{genetic algorithm} (GA) as an anytime heuristic to decouple these decision dimensions, rendering the massive search space computationally tractable and yielding effective schedules within practical runtime constraints.}



\revise{Specifically, each chromosome encodes a complete assignment of parallelism degrees $\{x_{c,k}\}$ for all cascades (Alg.~\ref{alg:ga_solver}, line~\textcolor{line}{5}). Its fitness is evaluated in a topology-agnostic manner by a fast list scheduler that heuristically determines launch times $\{s_c\}$ and the resulting makespan by prioritizing the critical path (lines~\textcolor{line}{6--7}). The population iteratively evolves through elitism, crossover, and mutation (line~\textcolor{line}{13}), enabling efficient exploration of the parallelism space and progressive refinement toward lower makespans.}




\subsection{Topology-aware Resource Mapper}\label{subsec:mapper}
Cascade-level parallelism planner treats GPUs as a homogeneous resource pool, which can result in initial allocations with substantial communication overhead. To mitigate this, Topology-aware resource mapper incorporates physical hardware topology during cascade placement, completing the planner’s logical schedule into a topology-aware, end-to-end execution plan for efficient T2V training.

\begin{figure*}[t]
      \centering
      \includegraphics[width=0.98\textwidth]{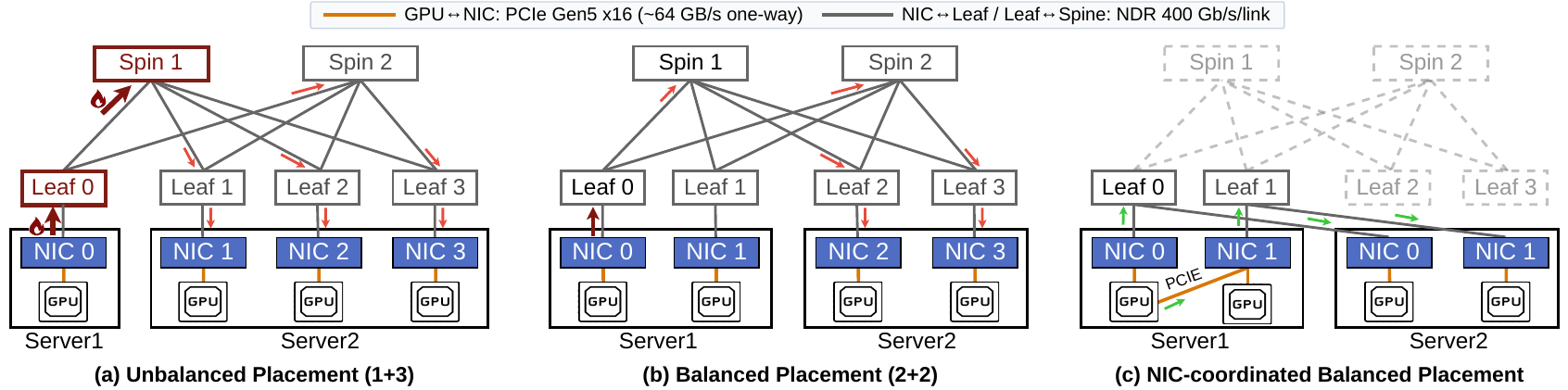}
      \caption{Comparison of three different resource placement strategies.}
      \label{fig:placement}
      \vspace{-1mm}
  \end{figure*}

\PHM{Intra-cascade Placement.}\label{subsubsec:intra_cascade}Topology-aware resource mapper prioritizes the communication patterns in a single cascade for resource assignments. This is achieved by using a \textit{four-level hierarchical prioritization scheme} that minimizes communication latency and prevents network bottlenecks.
\begin{itemize}[leftmargin=*]
\item\textit{Priority 1 (Within-Node Placement)}. The highest priority is to place an entire cascade within a single compute node. That is, if a node has enough available GPUs to satisfy its parallelism degree, Topology-aware resource mapper restricts placement to that node. Such localization ensures that collective communication operations (\eg, \texttt{All-Reduce}, \texttt{All-to-All}) are confined to high-bandwidth, low-latency intra-node interconnects such as NVLink.

\item\textit{Priority 2 (Balanced Cross-Node Placement)}. When within-node placement is infeasible, the priority is to distribute the cascade crossing nodes in a balanced manner. By balanced, we refer to an even distribution of inter-node communication that helps prevent localized congestion and reduces the likelihood of network hotspots. 

\PHM{Example 4.} For a cascade with a parallelism degree of $4$ (\texttt{SP=4}), the asymmetric \texttt{1+3} placement in~\autoref{fig:placement}(a) funnels cross-node traffic through a single shared
path, leading to severe contention on the PCIe bus, NIC, and the corresponding switch uplink/downlink resources (indicated by the fire symbol). In contrast, the symmetric \texttt{2+2} placement
in~\autoref{fig:placement}(b) balances communication across NICs, thereby mitigating congestion and improving performance.


\item\textit{Priority 3 (NIC-Coordinated Placement)}. Given a balanced inter-node placement (\eg, a \texttt{2+2} distribution), the third priority is to select specific GPUs that optimize data paths across the network fabric. This strategy leverages a common \textit{multi-rail cluster architecture} where same-numbered NICs across different nodes are connected to the same leaf switch.

\PHM{Example 5.}~\autoref{fig:placement}(c) illustrates the placement in which NIC indices are aligned across nodes ({\texttt{0,1}} on both), confining all inter-node communication to a single leaf switch (\eg, data from \texttt{GPU 0} (node A) to \texttt{GPU 1} (node B) travels via the local PCIe bus to \texttt{NIC 1} (node A), through leaf \texttt{switch 1}, reaches \texttt{NIC 1} (node B), and finally arrives at \texttt{GPU 1} (node B), completely avoiding the high-latency spine switch). The balanced placement in~\autoref{fig:placement}(b), however, assigns node A's GPUs to NICs {\texttt{0,1}} and node B's to {\texttt{2,3}}, forcing all inter-node traffic through the spine switch.


\item\textit{Priority 4 (Internal Communication-Aware Placement).} On modern GPU servers, latency between a GPU and a given NIC may vary due to traversal across PCIe switches or NUMA domains. In cases where multiple GPU combinations achieve equivalent NIC coordination, this strategy considers the topology of intra-node communication and selects the combination that minimizes the cumulative communication distance to the assigned NICs, thereby reducing intra-node latency and ensuring the most efficient path to the inter-node network.
\end{itemize}
\PHM{Inter-cascade Placement.}\label{subsubsec:inter_cascade}For interdependent cascades (e.g., a VAE passing its output to a DiT), we co-design placement with the data handoff mechanism introduced in Runtime executor (\S\ref{inter-cascade data transfer}) to enable efficient transfers. Topology-aware resource mapper follows a prioritized approach to achieve this: Its primary objective is to maximize the physical GPU overlap between the producer and consumer. A high degree of overlap allows Runtime executor to perform efficient in-place handoff, where shared GPUs act as a distributed cache and eliminate the need for explicit data movement. When such overlap is not feasible, the mapper optimizes a secondary objective\textemdash improving the efficiency of Runtime executor's handoff engine. It does so by co-locating the main rank of either the producer or the consumer with the central broker rank, thereby confining Gloo-based data transfers to high-speed intra-node interconnects and minimizing overhead for unavoidable data movement.

\begin{figure*}[!t]
    \centering
    \captionsetup{skip=2pt}
    \includegraphics[width=\textwidth]{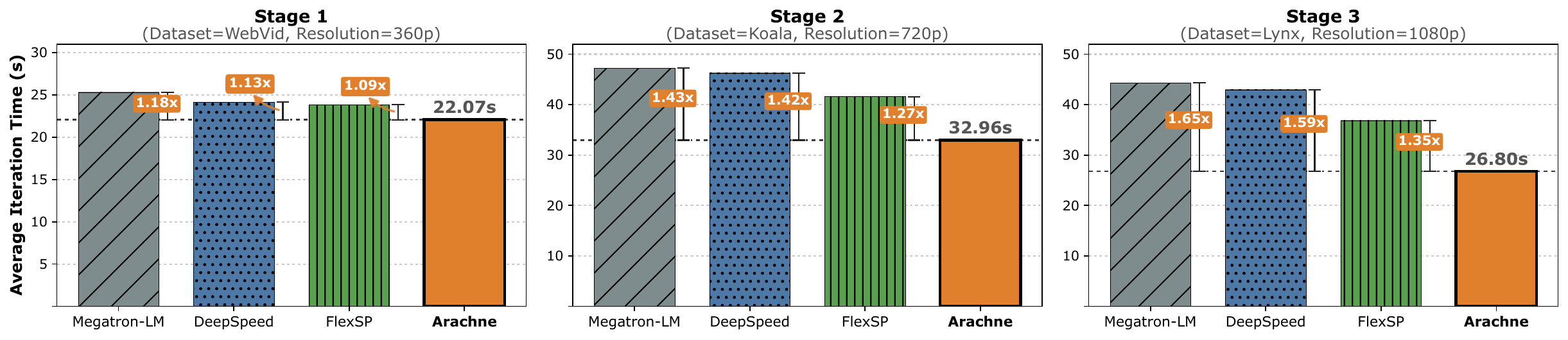}
    \caption{\textit{Average iteration time} (in seconds) across three training stages for the HunyuanVideo-13B model. The annotations above the \arachne{} bars indicate relative speedups compared to the baseline systems.}
    \vspace{-3mm}
    \label{fig:avg_iteartion_time_across_stage}
\end{figure*}

\subsection{Runtime Executor}\label{subsec:executor}
\noindent
Once it receives the optimized DAG from Topology-aware resource mapper, the primary role of Runtime executor is to translate this static plan into a live, dynamic execution on the cluster. It achieves this by treating each GPU as a dynamic resource that is rapidly re-tasked according to the DAG's spatio-temporal schedule. For instance, a single GPU might process a VAE cascade for one data sample and then be immediately re-assigned to a DiT cascade for another. To manage this fluidity, the Executor is responsible for two critical functions detailed below: orchestrating efficient inter-cascade data handoffs and resolving the complex challenge of heterogeneous gradient accumulation.



\PHM{Optimized Inter-cascade Data Handoff.}\label{inter-cascade data transfer}Runtime executor's key function in managing inter-cascade data dependencies (e.g., passing a latent tensor from a VAE to a DiT) is an adaptive data handoff mechanism. This mechanism avoids the prohibitive memory footprint of naively buffering tensors globally by leveraging the physical locality of the producer and consumer GPU groups, as determined by Topology-aware resource mapper.

The mechanism operates in two modes based on this locality. First, for overlapping GPU groups, it performs a highly efficient in-place handoff, where common GPUs act as a distributed cache. If the consumer group is a superset, a 
  localized broadcast from these caching GPUs efficiently updates the new peers. Second, for disjoint groups, a decoupled asynchronous engine allows the producer to ``fire-and-forget'' its output tensor via a stream-pipelined      
  pack-and-offload path, where tensors are packed on a secondary CUDA stream and copied to host memory asynchronously, then routed through locality-driven multi-broker nodes over a dedicated CPU-side Gloo channel that runs entirely
   in parallel with performance-critical NCCL collectives like \texttt{All-to-All}. Consumers further hide residual latency via lookahead prefetching of upcoming transfers. This adaptive strategy ensures inter-cascade data         
  dependencies are resolved with near-zero overhead on the critical training path.


\PHM{Heterogeneous Gradient Accumulation.}A key challenge arising from Runtime executor's dynamic cascade dispatch is heterogeneous gradient accumulation. The DAG schedule breaks the traditional one-to-one GPU-to-sample-gradient mapping, leading to a complex gradient landscape. For instance, while processing four data chunks $(A, B, C, D)$, one GPU subgroup might accumulate a composite gradient of $G_{A}+G_{B}$, a second subgroup could hold a different composite $G_{C}+G_{D}$, and a third might only contain a pure gradient $G_{A}$. This heterogeneity, where GPUs hold fundamentally different pre-summed gradient combinations, renders a conventional global \texttt{All-Reduce} mathematically incorrect, as it cannot disentangle the components.

To address this, Runtime executor introduces a two-phase                                                                                                                                                                             
  representative-based synchronization strategy. An exact
  minimum set cover selects the fewest representative GPUs                                                                                                                                                                             
  whose gradients collectively span all components. These                                                                                                                                                                              
  representatives first \texttt{All-Reduce} among themselves,                                                                                                                                                                          
  then each simultaneously broadcasts $G_{\text{final}}$ to 
  its non-representative peers, with both phases                                                                                                                                                                                       
  preferentially reusing existing execution groups. This
  ensures mathematically consistent gradient updates across                                                                                                                                                                            
  all GPUs with minimal synchronization overhead.

\section{Experiments}\label{sec: exp}

\subsection{Experimental Setup}

\PHB{Platform and Protocols.} All experiments run on a 64-GPU NVIDIA H100-80GB cluster (8 GPUs/node via NVLink, 8$\times$400\,Gbps InfiniBand NICs per node) using PyTorch 2.7.1 and CUDA 12.6. End-to-end results (\S\ref{subsec:end-to-end performance}) are reported on 16 GPUs, while scalability evaluations (\S\ref{sec:scalability}) use up to 64 GPUs. To ensure fair comparison, all systems use the same bucket configuration, adaptive batching, and activation checkpointing. We cap the maximum sequence length at 129 frames as a practical trade-off to maximize data coverage while preventing baseline OOM errors. Performance is averaged over 50 iterations following a 10-iteration warm-up.

\PHB{Baselines.} We compare \arachne{} with three state-of-the-art distributed training systems, namely \textbf{Megatron-LM}~\cite{Shoeybi2019MegatronLM} and \textbf{DeepSpeed}~\cite{Rajbhandari2020ZeRO}, as well as \textbf{FlexSP}~\cite{Wang2025FlexSP}, a representative LLM training system optimized for variable-length corpora. Here, Megatron-LM (DP with ZeRO-1 and CP) and DeepSpeed (ZeRO and Ulysses-style SP) serve as strong static baselines, utilizing fixed parallelism configurations (\eg, \texttt{DP=4}, \texttt{SP=4}) manually tuned for optimal performance. FlexSP adapts its SP degree per iteration based on bucket-level workload variations. In contrast to the baselines' static configurations, \arachne{} dynamically generates and executes an optimized fine-grained spatio-temporal plan for the cascades in each training iteration.

\begin{table}[!h]
\centering
\caption{Architectural details of the T2V models evaluated.}
\vspace{-1mm}
\setlength{\tabcolsep}{3pt}
\label{tab:model_architectures_video}
\begin{tabular}{lcccc}
    \toprule
    \textbf{T2V Models} & \textbf{\#Parameters}  & \textbf{\#Layers} & \textbf{\#Attn Heads} & \textbf{Hidden Dims} \\
    \midrule
    Wan2.1  & 1.3B    & 30 & 12 & 1536 \\
    CogVideoX & 5B   & 30 & 30 & 1920 \\
    HunyuanVideo & 13B & 60 & 24 & 3072 \\
    \bottomrule
\end{tabular}
\end{table}

\PHB{Workloads and Models.}\label{subsec:experimental workloads} To evaluate training performance under realistic conditions, we adopt the progressive curriculum from Open-Sora~\cite{Zheng2024OpenSora} (\S\ref{subsec:t2v_training_strategies}) across \textit{three stages}: Stage 1 on WebVid~\cite{WebVid} ($\sim$7M clips, 360p), Stage 2 on Koala~\cite{wang2024koala36mlargescalevideodataset} ($\sim$20M clips, 720p), and Stage 3 on an internal Lynx dataset ($\sim$10M clips, 1080p). We use three representative T2V models spanning 1.3B to 13B parameters to cover diverse architectures: \textbf{Wan2.1-1.3B}~\cite{Wang2025Wan} (standard self/cross-attention), \textbf{CogVideoX-5B}~\cite{Yang2025CogVideoX} (3D full attention), and \textbf{HunyuanVideo-13B}~\cite{Kong2024HunyuanVideo} (hybrid dual/single-stream). The architectural details are summarized in~\autoref{tab:model_architectures_video}.

\begin{figure}[!t]
  \centering
  \includegraphics[width=\linewidth]{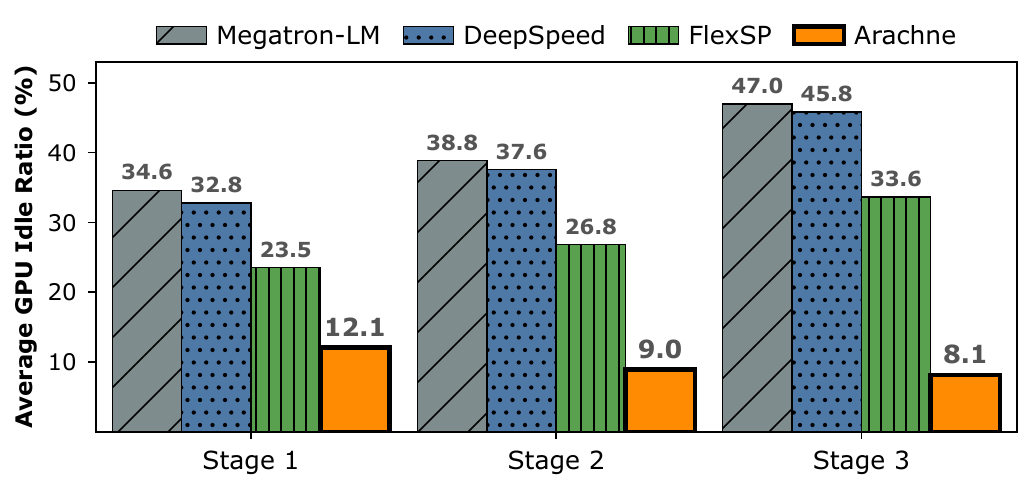}
  \caption{Average GPU idle ratio across three training stages.}
  \label{fig:avg_gpu_idle_ratio}
\end{figure}

\begin{figure*}[!t]
    \centering
    \includegraphics[width=\textwidth]{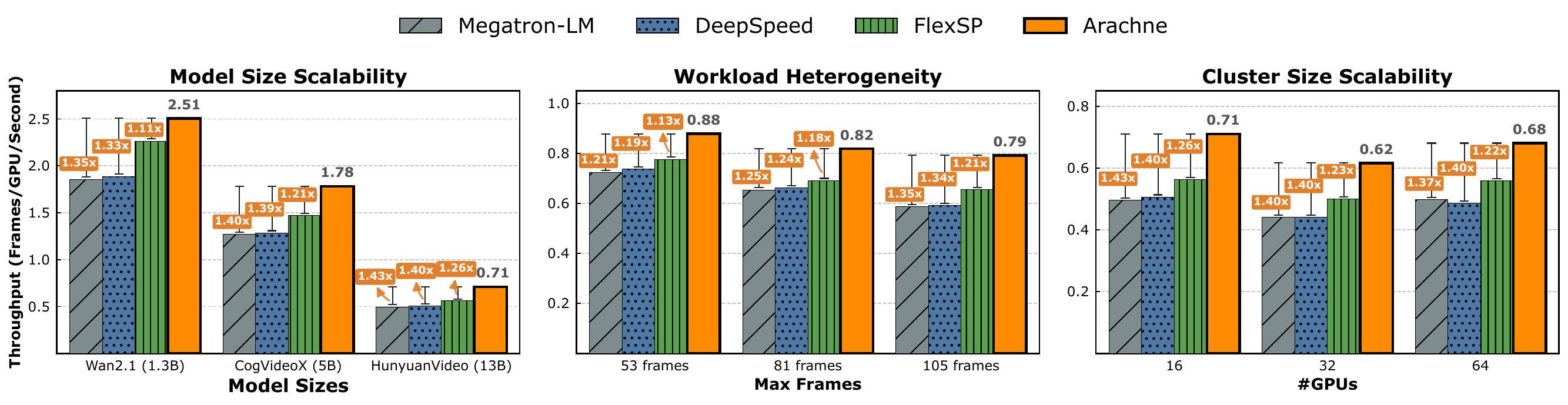}
\vspace{-2mm}
\caption{\textit{Throughput scalability evaluation} of \arachne{} under increasing training complexity, evaluated on HunyuanVideo-13B, across \textit{model size}, \textit{workload heterogeneity} (via larger maximum frame windows), and \textit{cluster size}.}
\vspace{-3mm}    
\label{fig:throught_comparison}
    
\end{figure*}

\subsection{End-to-End Performance}\label{subsec:end-to-end performance}
\PHB{Overall Results.} The main results for the training of HunyuanVideo are presented in~\autoref{fig:avg_iteartion_time_across_stage}. Overall, \arachne{} consistently outperforms all baselines across all three stages, with its performance advantage becoming increasingly pronounced from Stage 1 ($360$p) to Stage 3 ($1080$p). 
As data heterogeneity and computational density intensify at higher resolutions, the resulting workload imbalance exacerbates, thereby amplifying the benefits of \arachne{}'s fine-grained execution. \textbf{One remarkable outcome is that \arachne{} achieves peak speedups of 65\% over Megatron-LM, 59\% over DeepSpeed, and 35\% over FlexSP in Stage 3}.

\revise{
It is worth noting that the absolute average iteration time in Stage~3 (26.80s) is ``counterintuitively'' lower than that in Stage~2 (32.96s). This  observation actually stems from hardware memory constraints at $1080$p, which cap the maximum sequence length at 57 frames and thus reduce overall computation by skewing the workload distribution toward shorter sequences. Nevertheless, the computational disparity across buckets with different frame counts is substantially larger in Stage~3 ($1080$p), leading to more severe workload imbalance under static or coarse-grained parallelization. Notably, by effectively mitigating this amplified imbalance through fine-grained spatio-temporal orchestration, \arachne{} is able to capitalize on the increased heterogeneity in Stage~3, yielding its largest relative performance gains in this setting. }

We next analyze the average GPU idle ratio, defined as the fraction of total training time during which GPUs remain idle, in~\autoref{fig:avg_gpu_idle_ratio}. Across all three baselines, the idle ratio increases from Stage~1 to Stage~3, indicating that higher-resolution training exacerbates workload imbalance. FlexSP partially mitigates this trend by forming heterogeneous SP groups based on per-iteration sequence lengths, but its idle ratio remains substantial (rising to 33.6\% in Stage 3). In contrast, \arachne{} exhibits an inverse trend by dropping its idle ratio to 8.1\% across stages, confirming that fine-grained orchestration effectively converts growing data heterogeneity into richer scheduling opportunities rather than wasted cycles.

\subsection{Scalability Evaluation}\label{sec:scalability}
To evaluate the scalability of \arachne{}, we conduct experiments across three dimensions of large-scale training: \textit{computational intensity}, \textit{workload heterogeneity}, and \textit{cluster size}. Our exploration centers around the following three questions:

\begin{enumerate}[leftmargin=*, label={(\roman*)}]
  \item {How does \arachne{} perform as the training task becomes increasingly compute-intensive?}
  \item {To what extent can \arachne{} sustain efficiency under diverse and imbalanced workloads?}
  \item {How well does \arachne{} scale to larger clusters?}
\end{enumerate}

\PHM{Model-size Scalability.}We evaluate \arachne{}'s scalability with computational intensity by benchmarking across three T2V models of increasing size (Wan2.1-\underline{1.3B}, CogVideoX-\underline{5B}, and HunyuanVideo-\underline{13B}) under a fixed Stage~2 configuration, as shown in~\autoref{fig:throught_comparison} (left). The results reveal that \arachne{}'s performance advantage consistently amplifies with model scale: its throughput speedup over Megatron-LM grows from 35\% on the 1.3B model to 43\% on the 13B model. 
As larger models magnify the penalty of workload imbalance, both rigid static baselines and coarse-grained adaptations (\eg, FlexSP) suffer disproportionately. \arachne{}, conversely, effectively converts the resulting idle time into productive computation, demonstrating its ability to scale with computational intensity and underscoring its readiness for \textit{future, even larger models}.

\PHM{Workload-Heterogeneity Scalability.}Next, \autoref{fig:throught_comparison} (middle) evaluates scalability against workload heterogeneity by expanding the maximum frame window (\underline{53}, \underline{81}, and \underline{105} frames) for HunyuanVideo (Stage~2), sampling progressively wider, more imbalanced dataset portions (10.8\%, 46.9\%, and 70.4\% in~\autoref{fig:seq_dist_comparison}). As data diversity increases, \arachne{}'s throughput speedup over Megatron-LM surges from 21\% (53 frames) to 35\% (105 frames). Fundamentally, \arachne{} actively exploits this amplified workload imbalance as a richer scheduling space for its fine-grained orchestration, proving its robustness for \textit{diverse real-world datasets}.

\PHM{Cluster-size Scalability.}Finally, we evaluate cluster scalability by expanding the hardware from \underline{16} to \underline{32} and \underline{64} GPUs for HunyuanVideo under a fixed Stage~2 configuration in~\autoref{fig:throught_comparison} (right). Across all tested scales, \arachne{} maintains a stable and significant throughput advantage of 30\% to 40\% over both baselines. While scaling up heightens scheduling complexity with more concurrent cascades, \arachne{} effectively harnesses the simultaneously expanded spatial resource pool to unlock greater placement flexibility, thereby absorbing large-scale workload imbalances and highlighting its robust practicality for \textit{large-scale production training environments}.
In summary, experiments across these three dimensions demonstrate that \arachne{} exhibits a clear \textit{scaling trend}, consistently achieving efficiency gains with large-scale training scenarios. \textbf{This positions} \arachne{} \textbf{as a robust and future-ready system for large-scale training under ever-expanding computational and data demands.}


\subsection{Ablation Study}\label{subsec:ablation_study}
To validate the effectiveness of the resource placement strategies introduced in \arachne{} for minimizing communication overhead, we conduct an ablation study of Topology-aware resource mapper. All configurations are benchmarked on HunyuanVideo with three-stage training workloads from our end-to-end evaluation (\S\ref{subsec:end-to-end performance}).~\autoref{tab:ablation_mapper} presents the results.

\begin{table}[!t]
    \centering
    \caption{\textit{Ablation study} on HunyuanVideo-13B, showing the average iteration time (in seconds), with the corresponding slowdown shown in parentheses. ``w/o Mapper'' denotes removing Topology-aware resource mapper entirely, while ``w/o Inter-cascade'' and ``w/o Intra-cascade'' partially disable the inter- and intra-cascade placement optimization, respectively.}
    \label{tab:ablation_mapper}

    \setlength{\tabcolsep}{3pt}
    \begin{tabular}{l c c c}
    \toprule
    \textbf{Method} & \textbf{Stage 1} & \textbf{Stage 2} & \textbf{Stage 3} \\
    \midrule
    
    \arachne{} & \textbf{22.07} & \textbf{32.96} & \textbf{26.80} 
    \\
    \midrule
    w/o Inter-cascade  & \makecell[l]{22.98 (+4.13\%)} & \makecell[l]{34.29 (+4.04\%)} &  \makecell[l]{27.78 (+3.64\%)} \\
    w/o Intra-cascade & \makecell[l]{23.76 (+7.68\%)} & \makecell[l]{35.76 (+8.48\%)} & \makecell[l]{30.41 (\textbf{+13.67\%})} \\
    w/o Mapper  & \makecell[l]{23.89 (+8.26\%)} & \makecell[l]{37.09 (+12.53\%)} & \makecell[l]{31.89 (\textbf{+19.26\%})} \\
    \bottomrule
    \end{tabular}
\end{table}


    

The results show that disabling Topology-aware resource mapper significantly increases iteration times across all stages, with the most pronounced impact in Stage~3, which slows down by a substantial 19.26\%. A finer-grained analysis further reveals a clear hierarchy within the mapper's components. The intra-cascade placement strategy is the dominant contributor; disabling it in Stage~3 causes a 13.67\% slowdown, far greater than the 3.64\% slowdown from disabling the inter-cascade strategy. These findings validate our mapper's design, particularly its emphasis on optimizing intra-cascade communication. More fundamentally, they demonstrate that an optimized temporal plan is only truly effective when coupled with topology-aware spatial orchestration, confirming the necessity of our holistic spatio-temporal orchestration paradigm.

\subsection{Case Study}

To analyze how \arachne{} achieves its performance gains, we conduct a detailed case study of HunyuanVideo using Stage~2 workloads, combining execution-timeline visualization and workload-balancing analysis.

\begin{figure}[!h]
  \centering
  \includegraphics[width=.95\linewidth]{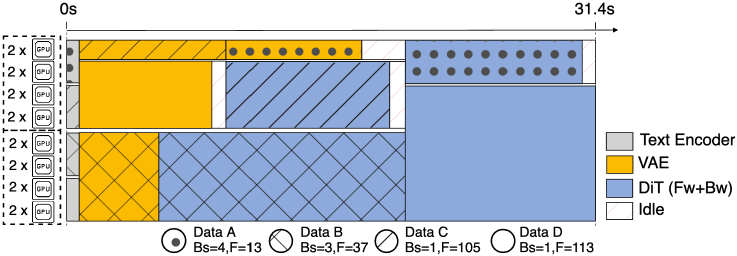}
  \caption{Execution timeline visualization in case study.}
  \label{fig:arachne_case}
\end{figure}

\YPHM{Execution Timeline Visualization}
We revisit the example introduced in~\autoref{fig:dynamic_parallelism_flex} to provide a before-and-after comparison against the execution schedule generated by \arachne{}. As shown in~\autoref{fig:arachne_case}, \arachne{} produces an optimized spatio-temporal execution by exercising fine-grained control over both temporal scheduling and spatial allocation of cascades, breaking the rigid constraints imposed by prior schedulers. Temporally, independent cascades are staggered to fill idle compute periods: shorter cascades from data $A$ and $B$ are dispatched to overlap with longer-running ones. Spatially, \arachne{} decouples cascades from fixed GPU sets to enable dynamic resource allocation. This behavior is exemplified by data $D$, whose text encoder, VAE, and DiT cascades are dynamically assigned 4 GPUs (\texttt{12--15}), 6 GPUs (\texttt{8--13}), and 12 GPUs (\texttt{0--11}), respectively. As a result, the makespan for this iteration is reduced to $31.4$s, representing a $30.5$\% reduction from the $45.2$s FlexSP baseline (as shown in ~\autoref{fig:dynamic_parallelism_flex}).



\begin{figure}[h]
    \centering
    \includegraphics[width=\linewidth]{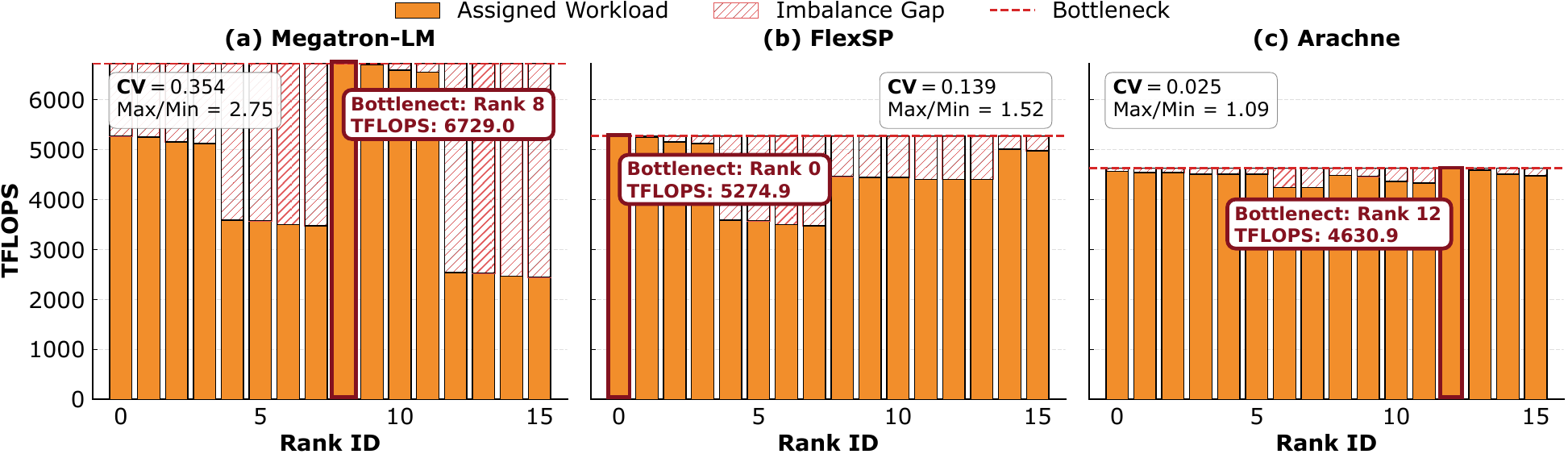}
    \caption{\textit{Per-rank TFLOPS distribution} in a training iteration. \textit{Hatched regions} show under-utilization relative to the bottleneck rank. Coefficient of Variation (CV) measures imbalance. Megatron-LM serves as the representative static baseline.}
    \label{fig:flops_dist}
\end{figure}

\begin{figure}[!h]
  \centering
  \includegraphics[width=0.8\linewidth]{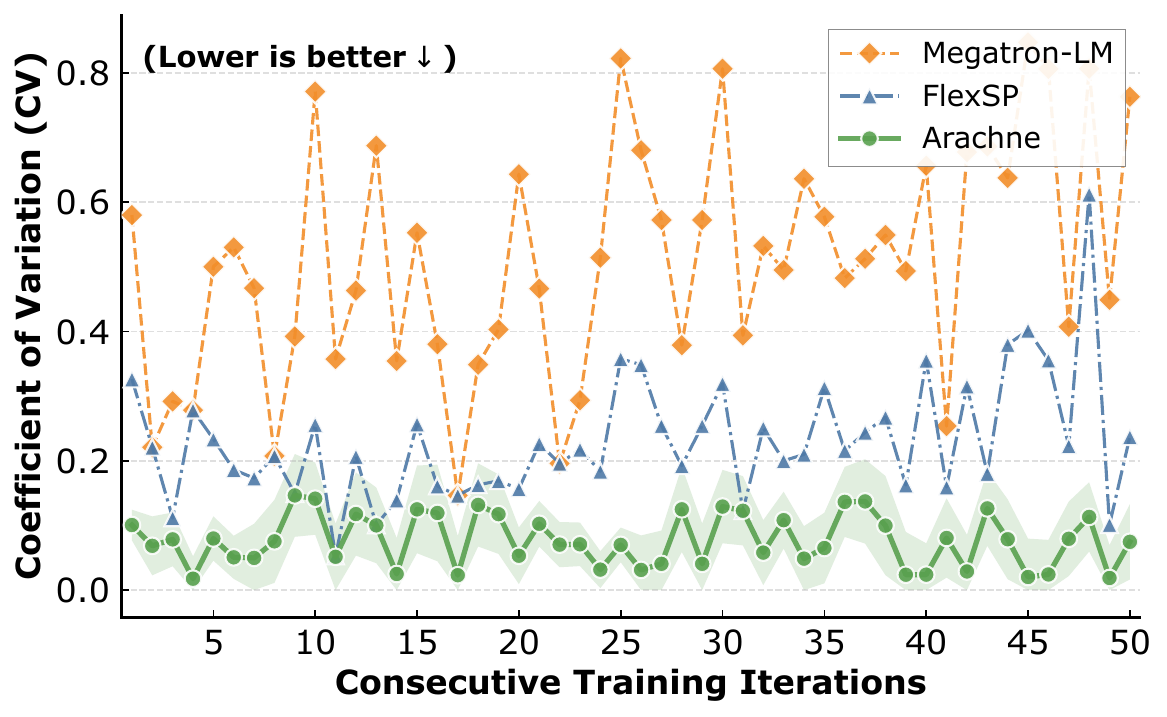}
  \vspace{-1mm}
  \caption{\textit{Temporal stability} measured by CV of per-rank TFLOPS over $50$ consecutive training iterations.}
  \label{fig:cv_stability}
\end{figure}

\YPHM{Workload Balancing Analysis.}\revise{
To attribute the observed performance gains to improved workload balancing, we analyze both the \textit{spatial distribution} and \textit{temporal stability} of per-rank computation. 
~\autoref{fig:flops_dist} presents the per-rank TFLOPS distribution within a single training iteration. Compared to all baselines, \arachne{} achieves a more balanced workload across GPU ranks, as evidenced by both a lower CV and smaller imbalance gaps.
~\autoref{fig:cv_stability}, on the other hand, measures the CV of per-rank TFLOPS over 50 consecutive training iterations. \arachne{} consistently maintains a low CV, remaining around 0.10 across iterations, whereas the baselines exhibit substantially higher variance over time. 
Taken together, these results demonstrate that \arachne{}’s fine-grained spatio-temporal orchestration yields a workload allocation that is both spatially balanced and temporally stable, underpinning its sustained training throughput improvements.
}

\section{Conclusion}\label{sec:conclusion}

We presented \arachne{}, a novel distributed training framework for efficient T2V model training, based on the fine-grained spatio-temporal orchestration paradigm. Decomposing the training workload into fine-grained cascades, \arachne{} optimized their physical placement and execution timeline, addressing the severe workload imbalance caused by data heterogeneity.  The comprehensive evaluation on leading T2V models showed that \arachne{} achieved up to 65\% reduction in iteration time compared to state-of-the-art training frameworks. More importantly, it revealed that the performance advantage of this approach scales with both model complexity and workload heterogeneity, confirming its robustness and practicality as a solution for large-scale generative model training in the future.


\bibliographystyle{plain}
\bibliography{software}

\end{document}